\documentstyle[aps,prl,epsfig,float,twocolumn]{revtex}
\begin{document}
\twocolumn[\hsize\textwidth\columnwidth\hsize\csname
@twocolumnfalse\endcsname

\title{Ballistic side jump motion of electrons and holes in semiconductor
quantum wells}

\author{John Schliemann}

\address{Institute for Theoretical Physics, University of 
Regensburg, D-93040 Regensburg, Germany}

\date{\today}

\maketitle

\begin{abstract}
We investigate the ballistic motion of electrons and holes in III-V 
semiconductor quantum wells with spin-orbit coupling 
and a homogeneous  in-plane electric field. As a result of a non-perturbative
treatment of both of these influences, particle wave packets undergo a 
pronounced side jump perpendicular to the field direction.
For wave packets of sufficient width
the amplitude of this motion can be estimated analytically and increases with
decreasing field strength. We discuss the scaling behavior of the effect and
some if its experimental implications.
\end{abstract}
\vskip2pc]

Effects of spin-orbit coupling in semiconductors have become a particularly
lively direction of solid-state research. Most of the activities here are
embedded in the emerging field of spintronics. An early paradigm of this area
is the spin field-effect transistor proposed by Datta and Das over fifteen 
years ago \cite{Datta90}. In this device proposal, electrons are injected 
from a 
spin-polarized source electrode into a quantum well in which the dominant
contribution to spin-orbit interaction is given by the Rashba coupling,
\begin{equation}
{\cal H}_{R}=\frac{\alpha}{\hbar}\left(p_{x}\sigma^{y}-p_{y}\sigma^{x}\right).
\label{rashba}
\end{equation}
Here $\vec p$ is the momentum of the electron confined in a
two-dimensional geometry, and $\vec\sigma$ the vector of Pauli
matrices. The Rashba coefficient $\alpha$ is tunable by an electric gate across the
well and can therefore be varied experimentally
\cite{Rashba60}. As a result, in the 
course of the electron's ballistic motion through the quantum well its spin
undergoes a controlled rotation which can be detected by another spin-polarized
drain electrode. To enable the injection of electrons into this device, a
certain voltage has to be applied between source and drain electrode, leading
to a possibly small but nevertheless finite in-plane electric field in the
quantum well. In the present letter we investigate the effect of such an
electric field in combination with spin-orbit coupling. As we shall see below,
the interplay between spin-orbit coupling and an in-plane electric field
adds an important feature to the ballistic electron dynamics: The electron
performs a side-jump motion perpendicular to the direction of the 
electric field. The field here is assumed to be homogeneous, in contrast to 
side jump motion upon scattering on impurities which is discussed intensively 
in circumstances of the anomalous Hall effect \cite{Sinova04a}.

Moreover, as 
it was pointed out recently, spin-orbit coupling of the type (\ref{rashba})
does not only lead to a rotation of the spin, but has also
an effect on the orbital degree of freedom: the electron performs a
{\em zitterbewegung}, i.e. an oscillatory motion superimposed on the
nonrelativistic dynamics of the particle wave packet 
\cite{Schliemann05a}. 
In the absence of an electric field the
single-particle Hamiltonian reads ${\cal H}=\frac{\vec p^{2}}{2m}+{\cal H}_{R}$
where $m$ is the effective band mass. In this case a fully analytical 
description of the {\em zitterbewegung} can be given in terms of the
time-dependent position operator
$\vec r_{H}(t)=\exp({i{\cal H}t/\hbar})\vec r(0)\exp({-i{\cal H}t/\hbar})$
in the Heisenberg picture. As we shall see, the side jump phenomena
to be discussed below are accompanied by {\em zitterbewegung} of this type.

Another important recent development regarding spin dynamics in semiconductors
is the prediction of the intrinsic spin Hall effect  
\cite{Murakami03,Sinova04b,Schliemann05b,Schliemann06}. This effect is again
a consequence of spin-orbit coupling and amounts in a spin current (as opposed
to a charge current) driven by a perpendicular electric field. Therefore, this 
situation is similar to the Datta-Das transistor with the electric field being
provided by the source-drain voltage. Here spin orbit coupling
of the Rashba type has the peculiarity that 
the spin Hall conductivity vanishes in an infinite system in the
presence of any spin relaxation mechanism \cite{note1}. However, this 
conclusion holds only in the thermodynamic limit, and for a device 
of mesoscopic size and sufficient carrier mobility, spin Hall transport
signaled by spin accumulation at the sample edges should be considered
as possible.

Let us consider an electron in an n-doped quantum well being subject to 
Rashba spin-orbit coupling and a constant in-plan electric force
$\vec F$, i.e. the single-particle Hamiltonian reads
${\cal H}=\frac{\vec p^{2}}{2m}+{\cal H}_{R}-\vec F\vec r$. Thus, the
time-dependent momentum operator is given by 
$\vec p_{H}(t)=\vec p(0)+\vec Ft$, while the spin operators fulfill the
equations
\begin{eqnarray}
\frac{d}{dt}\sigma^{x}_{H}(t)& = & \frac{2\alpha}{\hbar^{2}}p^{x}_{H}(t)\sigma^{z}_{H}(t)\,,
\label{sigmax}\\
\frac{d}{dt}\sigma^{y}_{H}(t)& = & \frac{2\alpha}{\hbar^{2}}p^{y}_{H}(t)\sigma^{z}_{H}(t)\,,
\label{sigmay}\\
\frac{d}{dt}\sigma^{z}_{H}(t)& = & -\frac{2\alpha}{\hbar^{2}}
\left[p^{x}_{H}(t)\sigma^{x}_{H}(t)+p^{y}_{H}(t)\sigma^{y}_{H}(t)\right]
\label{sigmaz}
\end{eqnarray}
with $\vec\sigma_{H}(0)=\vec\sigma$. Note that the position operator $\vec r$ does not
occur in the above equations, and therefore the time-dependent
spin operators  $\vec\sigma_{H}(t)$ can be formulated as a function of the
zero-time Schr\"odinger operators $\vec p$, $\vec\sigma$ only. In particular, when
the operators $\vec\sigma_{H}(t)$ are applied on eigenstates of the 
momentum $\vec p=\vec p(0)$, the quantities $\vec p_{H}(t)=\vec p(0)+\vec Ft$
become real numbers, rendering Eqs.~(\ref{sigmax})-(\ref{sigmaz}) as an
system of ordinary differential equations for the $2\times 2$-matrices
$\vec\sigma_{H}(t)$. Unfortunately an explicit analytical solution
of these equations for general directions of the force $\vec F$ and the
initial momentum $\vec p(0)$ does not seem to be possible. However, such a
solution can be given if $\vec F$ and $\vec p(0)$ are collinear.
Since the Rashba Hamiltonian (\ref{rashba}) is invariant under rotations
of spin and momentum in the xy-plane, 
we can choose, without loss of generality, $\vec F$ and $\vec p(0)$ to point
along the $x$-direction where we find
\begin{eqnarray}
\sigma^{x}_{H}(t)& = &\sigma^{x}
\cos\left(\frac{2\alpha}{\hbar^{2}}\left(p^{x}t+\frac{1}{2}Ft^{2}\right)\right)\nonumber\\
 & + & \sigma^{z}
\sin\left(\frac{2\alpha}{\hbar^{2}}\left(p^{x}t+\frac{1}{2}Ft^{2}\right)\right)\,,\\
\sigma^{y}_{H}(t)& = &\sigma^{y}\,,\\
\sigma^{z}_{H}(t)& = &-\sigma^{x}
\sin\left(\frac{2\alpha}{\hbar^{2}}\left(p^{x}t+\frac{1}{2}Ft^{2}\right)\right)\nonumber\\
 & + & \sigma^{z}
\cos\left(\frac{2\alpha}{\hbar^{2}}\left(p^{x}t+\frac{1}{2}Ft^{2}\right)\right)\,.
\end{eqnarray}
Note that the argument of the trigonometric functions is the integral
$\int_{0}^{t}dt'p^{x}_{H}(t')$, generalizing the situation without an external 
field and therefore constant momentum.

\begin{center}
\begin{figure}
\epsfig{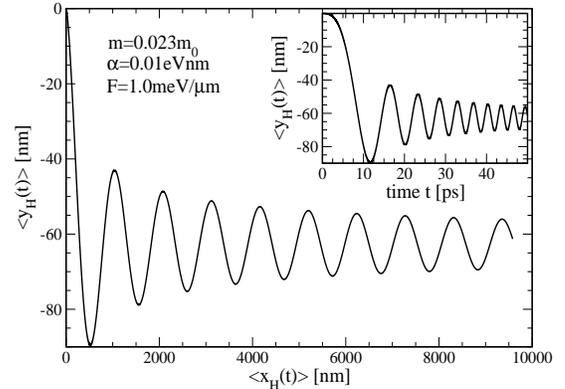}
\caption{Main panel: real-space motion $\langle\vec r_{H}(t)\rangle$ for a plane
wave with $\vec p(0)=0$. The inset shows $\langle y_{H}(t)\rangle$ as a 
function of time $t$.}
\label{fig1}
\end{figure}
\end{center}

The time-dependent position operators $\vec r_{H}(t)$ fulfill the equations
\begin{eqnarray}
\frac{d}{dt}x_{H}(t) & = & \frac{p^{x}_{H}(t)}{m}+\frac{\alpha}{\hbar}\sigma^{y}_{H}(t)\,,\\
\frac{d}{dt}y_{H}(t) & = & \frac{p^{y}_{H}(t)}{m}-\frac{\alpha}{\hbar}\sigma^{x}_{H}(t)\,.
\end{eqnarray}
Let us now consider the expectation values $\langle\vec r_{H}(t)\rangle$ for an initial
state with momentum $\vec p\parallel\vec F$ along the $x$-direction, spin
in positive $z$-direction and $\langle\vec r_{H}(0)\rangle=0$. Here we have
$\langle x_{H}(t)\rangle=(p^{x}t+\frac{1}{2}Ft^{2})/m$, and the $y$-component reads
\begin{equation}
\langle y_{H}(t)\rangle=-\frac{\alpha}{\hbar}\int_{0}^{t}dt'
\sin\left(\frac{2\alpha}{\hbar^{2}}\left(p^{x}t'+\frac{1}{2}Ft'^{2}\right)\right)\\.
\label{y1}
\end{equation}
If the initial spin direction is reversed, $\langle y_{H}(t)\rangle$ changes sign.
The case of vanishing initial momentum $\vec p(0)=0$ is particularly
interesting. Here the physical problem contains only two length scales
which are conveniently chosen as 
the Rashba length $\lambda_{R}=\hbar^{2}/m\alpha^{2}$ and the field-dependent length
$l_{F}=\sqrt{\alpha/F}$. The latter quantity
determines the amplitude of the side jump motion
to be discussed now. 
Figure \ref{fig1} shows a plot of $\langle y_{H}(t)\rangle$ for $\vec p(0)=0$, 
$F=1.0{\rm meV/}\mu{\rm m}$,
and a band mass $m=0.023m_{0}$, corresponding to the conduction band mass on InAs,
where $m_{0}$ is the bare electron mass. The Rashba parameter has been fixed to
$\alpha=0.01{\rm eVnm}$ which is a realistic value for InAs \cite{Schliemann06}. 
The main panel shows shows the a real-space plot of $\langle\vec r_{H}(t)\rangle$ with the time
$t$ ranging from zero to $t=50{\rm ps}$. One can
clearly distinguish two phases of the electron motion: In a first phase with
$\langle x_{H}(t)\rangle\lesssim 200{\rm nm}$, the transverse expectation value $\langle y_{H}(t)\rangle$
starts at $\langle y_{H}(0)\rangle=0$ and undergoes a pronounced monotonic side jump motion
to  $|\langle y_{H}(t)\rangle|\approx 60{\rm nm}$. In a following second phase
$\langle y_{H}(t)\rangle$ performs a {\em zitterbewegung}, 
i.e. an oscillatory motion around this value with its
amplitude decreasing with increasing longitudinal distance $\langle x_{H}(t)\rangle$.
The period of the {\em zitterbewegung} as a function of $\langle x_{H}(t)\rangle$ is
approximately given by $\pi\lambda_{R}\approx 1040{\rm nm}$ \cite{Schliemann05a}.
The inset of figure \ref{fig1} shows the same data for $\langle y_{H}(t)\rangle$ but as a 
function of time $t$. 
After a time interval of $t=50{\rm ps}$ the electron 
wave number is $k^{x}(t)=p^{x}(t)/ \hbar=Ft/ \hbar<0.08{\rm nm^{-1}}$ which is still 
close to the $\Gamma$-point in III-V semiconductors. Thus, the assumptions 
underlying the effective-mass approximation and the Rashba Hamiltonian 
(\ref{rashba}) are perfectly valid. 

As seen from figure \ref{fig1}, the magnitude of the side jump can be obtained
by performing the formal limit
\begin{eqnarray}
\lim_{t\to\infty}\langle y_{H}(t)\rangle & = & -\frac{\alpha}{\hbar}\int_{0}^{\infty}dt
\sin\left(\frac{\alpha}{\hbar^{2}}Ft^{2}\right)\\
 & = & -l_{F}\sqrt{\frac{\pi}{8}}\,,
\label{limit1}
\end{eqnarray}
where we have used the integral 
$\int_{0}^{\infty}dt\sin(t^{2})=\sqrt{\pi/8}\approx 0.62$. For the above parameters we have
$l_{F}=100{\rm nm}$ corresponding to a  side jump of about $62{\rm nm}$, in 
perfect agreement with the numerical plot. Note that according to 
Eq.~(\ref{y1}) the band mass
$m$ does not enter the function $\langle y_{H}(t)\rangle$, but is inversely
proportional to $\langle x_{H}(t)\rangle$. Thus, changing the effective mass as
$m\mapsto m'$ amounts in a
rescaling of the abscissa in the main panel of figure \ref{fig1}
by a factor of $m/m'$. Thus the above considerations relating to InAs are
trivially extended to other materials, say GaAs. In the latter case
the Rashba coupling is typically somewhat smaller which can be compensated by
a smaller in-plane field $F$, resulting the the same length scale $l_{F}$, i.e.
the same magnitude of side jump.

In the above discussion we have concentrated on plane-wave
eigenstates of the momentum. However, a localized  electron in a quantum well 
is not described realistically by a single plane wave but rather a 
superposition of them. We therefore consider a Gaussian wave packet,
\begin{equation}
\langle\vec r|\psi\rangle=\frac{1}{2\pi}\frac{d}{\sqrt{\pi}}
\int d^{2}k\,e^{-\frac{1}{2}d^{2}\left(\vec k-\vec k_{0}\right)^{2}}
e^{i\vec k\vec r}
\left(
\begin{array}{c}
1 \\ 0
\end{array}
\right)\,,
\end{equation}
where the initial spin direction is again along the $z$-axis.
We numerically solve for the dynamics of such a wave packet by superimposing
plane-wave solutions whose initial momenta have been chosen from
a grid centered around $\vec k_{0}$
in reciprocal space. Because the in-plane field $\vec F$ is in
general not collinear with such initial momenta, these plane-wave solutions
have also to be generated numerically. Moreover, in such a procedure
necessarily requires a wave vector cutoff. In all numerical simulations
to be presented below we have carefully checked that the data is well-converged
with respect to grid spacing and wave vector cutoff.   

\begin{center}
\begin{figure}
\epsfig{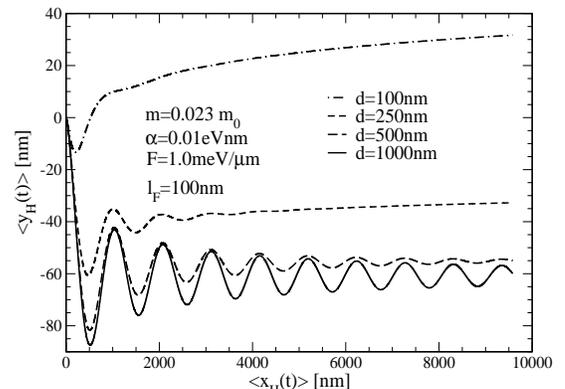}
\caption{Real-space motion $\langle\vec r_{H}(t)\rangle$ for electron wave packets
with zero initial group velocity and different width $d$.}
\label{fig2}
\end{figure}
\end{center}

For simplicity, let us 
concentrate again on the case of zero initial group wave number, $\vec k_{0}=0$.
Then the width $d$ of the wave packet is the only new length entering the 
problem. Figure \ref{fig2} shows numerical simulations of the real space 
dynamics for different wave packet width $d$ with otherwise the same 
parameters as before, $l_{F}=100{\rm nm}$. As seen from the figure,
the dynamics of a single plane wave discussed in Eqs.~(\ref{y1}),
(\ref{limit1}) is a good approximation to the
wave-packet motion if $d\gg l_{F}$. In fact, for $d=1000{\rm nm}=10l_{F}$ the dynamics
of the wave packet is very close to the plane-wave result shown in 
figure \ref{fig1}, which continues to be a good approximation
up to $d\approx 500{\rm nm}=5l_{F}$, whereas for smaller $d$ the nature of the
motion clearly changes. The criterion $d\gg l_{F}$ follows from the observation
that Eqs.~(\ref{sigmax})-(\ref{sigmaz}) are invariant under the rescaling
$\vec F\mapsto q\vec F$, $\vec p\mapsto \sqrt{q}\vec p$, $t\mapsto t/ \sqrt{q}$ with $q$ being
a real parameter. Therefore, in order to ensure 
invariance, the width $d$ of the wave packet has to scale as $d/ \sqrt{F}$
which uniquely determines $l_{F}=\sqrt{\alpha/F}$ as the relevant length scale.
Note also that Eqs.~(\ref{sigmax})-(\ref{sigmaz}) are also invariant under
the replacements $(p^{i}_{H}(t),\sigma^{i}_{H}(t))\mapsto(-p^{i}_{H}(t),-\sigma^{i}_{H}(t))$, $i\in\{x,y\}$.
As a consequence, for the above situation of a wave packet symmetric around
its group wave vector $\vec k_{0}=0$ and the force $\vec F$ pointing along the
$x$-direction, we have the scaling behavior $\langle y_{H}(t)\rangle\mapsto\langle y_{H}(t)\rangle/ \sqrt{q}$,
whereas $\langle x_{H}(t)\rangle$ remains unchanged. The latter conclusions hold for any 
symmetric wave packet, independently of its form and width; in particular,
the magnitude of the side jump scales always the same way as $l_{F}$.
Analogous findings hold for the rescaling
$\alpha\mapsto q\alpha$, $\vec p\mapsto \vec p/ \sqrt{q}$, $t\mapsto t/ \sqrt{q}$ leading to
$\langle x_{H}(t)\rangle\mapsto\langle x_{H}(t)\rangle/ \sqrt{q}$, $\langle y_{H}(t)\rangle\mapsto\sqrt{q}\langle y_{H}(t)\rangle$.
Therefore, figure \ref{fig2}
is of universal character since data for other values of $\alpha$, $F$ can be
obtained by a simple rescaling of the coordinate axes.
 
So far we have concentrated on the Rashba Hamiltonian (\ref{rashba})
as an effective contribution to spin orbit coupling. The situation is changed
only marginally if the Rashba coupling is replaced with the Dresselhaus term
\cite{Dresselhaus55} in its linear approximation \cite{Dyakonov86}.
When both terms are present an analytical solution for $\vec\sigma_{H}(t)$ does
not appear to be possible even for initial conditions with
$\vec p(0)\parallel \vec F$. An exception is the case when the Dresselhaus and the 
Rashba term are equal in magnitude: Here side jump and {\em zitterbewegung}
are absent because of an additional conserved quantity arising at this
high-symmetry point \cite{Schliemann03}.

\begin{center}
\begin{figure}
\epsfig{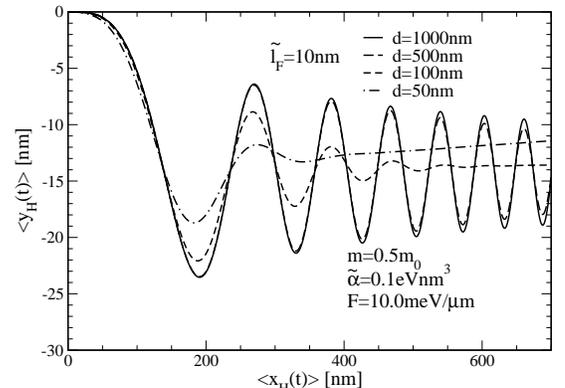}
\caption{Real-space motion \protect{$\langle\vec r_{H}(t)\rangle$} for hole wave packets
with zero initial group velocity and different width $d$.}
\label{fig3}
\end{figure}
\end{center}

Let us now turn to p-doped quantum wells of III-V semiconductors. Here the
low-energy physics is dominated by the heavy holes, and the spin-orbit
interaction term analogous to the Rashba coupling on electron spins reads
\cite{Winkler00}
\begin{equation}
\tilde{\cal H}_{R}=i\frac{\tilde\alpha}{2\hbar^{3}}\left(p_{-}^{3}\sigma_{+}-p_{+}^{3}\sigma_{-}\right)\,,
\label{defham}
\end{equation}
using the notations $p_{\pm}=p_{x}\pm ip_{y}$, $\sigma_{\pm}=\sigma^{x}\pm i\sigma^{y}$.
The  Pauli matrices operate on the total angular momentum states
with spin projection $\pm 3/2$ along the growth direction chosen as then $z$-axis,and $\tilde\alpha$ is the spin-orbit coupling coefficient. This Hamiltonian
contains two length scales which are, similarly to the
previous case of conduction-band electrons, aptly chosen to be
the field-dependent length  $\tilde l_{F}=(\tilde\alpha/F)^{1/4}$  and  the Rashba 
length $\tilde\lambda_{R}=m\tilde\alpha^{2}/ \hbar^{2}$
where $m$ is the effective heavy-hole band mass. 
Moreover, an analytical plane-wave solution 
can again be found if the initial momentum $\vec p(0)$ is collinear 
with the in-plane
force $\vec F$. Choosing this direction, again without loss of generality, to
lie along the $x$-axis, one finds
\begin{eqnarray}
\sigma^{x}_{H}(t)& = &\sigma^{x}
\cos\left(\frac{2\tilde\alpha}{\hbar^{4}}\int_{0}^{t}dt'\left(p^{x}+Ft'\right)^{3}\right)
\nonumber\\
 & - & \sigma^{z}
\sin\left(\frac{2\tilde\alpha}{\hbar^{4}}\int_{0}^{t}dt'\left(p^{x}+Ft'\right)^{3}\right)\,,\\
\sigma^{y}_{H}(t)& = &\sigma^{y}\,,\\
\sigma^{z}_{H}(t)& = &\sigma^{x}
\sin\left(\frac{2\tilde\alpha}{\hbar^{4}}\int_{0}^{t}dt'\left(p^{x}+Ft'\right)^{3}\right)
\nonumber\\
 & + & \sigma^{z}
\cos\left(\frac{2\tilde\alpha}{\hbar^{4}}\int_{0}^{t}dt'\left(p^{x}+Ft'\right)^{3}\right)\,,
\end{eqnarray}
where the remaining integral is elementary. 
For a plane-wave state with $\vec p(0)=0$, $\langle\vec r_{H}(0)\rangle=0$ one finds again
$\langle x_{H}(t)\rangle=Ft^{2}/2m$ and 
\begin{equation}
\langle y_{H}(t)\rangle=-\frac{3\tilde\alpha}{\hbar^{3}}\int_{0}^{t}dt'(Ft')^{2}
\sin\left(\frac{\alpha}{2\hbar^{4}}F^{3}t'^{4}\right)\\.
\label{y2}
\end{equation}
The magnitude of the side jump is given by the limit
\begin{eqnarray}
\lim_{t\to\infty}\langle y_{H}(t)\rangle & = & -3\cdot 2^{3/4}\left(\frac{\tilde\alpha}{F}\right)^{1/4}\int_{0}^{\infty}dxx^{2}
\sin\left(x^{4}\right)\nonumber\\
 & \approx & -1.428\cdot\tilde l_{F}\,.
\label{limit2}
\end{eqnarray}
Thus, the length scale $\tilde l_{F}$ plays a similar role as $l_{F}$
for conduction-band electrons. Moreover, analogously to this previous case,
the behavior of plane-wave states is mimicked by proper wave packets
provided $d\gg\tilde l_{F}$, a conclusion following from similar scaling 
arguments as before. The situation is illustrated in figure \ref{fig3}
with numerical simulations for values for $m$, $\tilde\alpha$ 
appropriate for heavy holes in p-doped GaAs quantum wells and 
$\tilde l_{F}=10 {\rm nm}$.
For wave packet width up to $d\approx5\tilde l_{F}$
the magnitude of the side jump is quantitatively well described by
Eq.~(\ref{limit2}), whereas for smaller values of $d$ the nature of the 
dynamics changes.

In summary, the pronounced side jump found in the dynamics of 
electrons and holes in semiconductor quantum wells
is the non-perturbative effect of the interplay of spin-orbit coupling and a
homogeneous electric field. The magnitude of the side jump motion is
governed by the field-dependent length scales $l_{F}=\sqrt{\alpha/F}$ and
$\tilde l_{F}=(\tilde\alpha/F)^{1/4}$ for electrons and holes, respectively, provided
that the width $d$ of the particle wave packet being initially at rest 
fulfills 
$d\gtrsim 5(l_{F},\tilde l_{F})$. A possible scenario for the experimental 
detection of the side jump effect are two opposite quantum point contacts of 
width $d$ whose centers are displaced by $\Delta\approx 0.2d$. The conductivity of 
this setup should be maximal if the experimentally tunable parameters
$\alpha$ ($\tilde\alpha$) and $F$ are adjusted such that 
$l_{F}(\tilde l_{F})\approx\Delta$.

Another interesting question is the connection between the ballistic side jump 
motion and the intrinsic spin Hall effect in diffusive systems. In a heuristic
but appealing picture one can interpret spin hall transport as iterated
side jumps occurring during the ballistic motion between scattering events.
In fact a theoretical synthesis of extrinsic and intrinsic mechanisms of spin 
Hall transport \cite{Schliemann06} was put forward in 
Ref.~\cite{Hankiewicz06} where the side jump contribution to the spin Hall
conductivity was found to be independent of disorder and particle
interactions.

I thank I. Adagideli and D. Weiss for useful discussions. 
This work was supported by the SFB 689  
``Spin Phenomena in reduced Dimensions''.



\begin{references}

\bibitem{Datta90}
S. Datta and B. Das, Appl. Phys. Lett. {\bf 56}, 665 (1990).

\bibitem{Rashba60} E.~I. Rashba, Fiz. Tverd. Tela (Leningrad) \textbf{2},
1224 (1960) (Sov. Phys. Solid State \textbf{2}, 1109 (1960)); Y.~A. Bychkov
and E.~I. Rashba, J. Phys. C \textbf{17}, 6039 (1984).

\bibitem{Sinova04a}
For a historical overview see J. Sinova, T. Jungwirth, and J. Cerne,
Int. J. Mod . Phys. B {\bf 18}, 1083 (2004).

\bibitem{Schliemann05a}
J. Schliemann, D. Loss, and R.~M. Westervelt,
Phys. Rev. Lett. {\bf 94}, 206801 (2005);
Phys. Rev. B {\bf 73}, 085323 (2006). 

\bibitem{Murakami03}
S. Murakami, N. Nagaosa, and S.~C. Zhang, Science {\bf 301}, 1348 (2003).

\bibitem{Sinova04b}
J. Sinova, D. Culcer, Q. Niu, N.~A. Sinitsyn, T. Jungwirth, 
and A.~H. MacDonald, Phys. Rev. Lett. {\bf 92}, 126603 (2004).

\bibitem{Schliemann05b}
J. Schliemann and D. Loss, Phys. Rev. B {\bf 71}, 085308 (2005).

\bibitem{Schliemann06}
J. Schliemann, Int. J. Mod. Phys. B {\bf 20}, 1015 (2006).

\bibitem{note1}
This  result was obtained first by J.~I. Inoue, G.~E.~W. Bauer, 
and L.~W. Molenkamp, Phys. Rev. B {\bf 70}, 041303 (2004), studying spin Hall
transport in the presence of static impurities within a diagrammatic 
expansion. For further developments and generalizations of this conclusion
see Ref.~\cite{Schliemann06}.

\bibitem{Dresselhaus55}
G. Dresselhaus, Phys. Rev. {\bf 100}, 580 (1955). 

\bibitem{Dyakonov86}
M.~I. Dyakonov and V.~Y. Kachorovskii,
Sov. Phys. Semicond. {\bf 20}, 110 (1986);
G. Bastard and R. Ferreira, Surf. Science {\bf 267}, 335 (1992).

\bibitem{Schliemann03}
J. Schliemann, J.~C. Egues, and D. Loss, 
Phys. Rev. Lett. {\bf 90}, 146801 (2003).

\bibitem{Winkler00}
R. Winkler, Phys. Rev. B {\bf 62}, 4245 (2000);
L.~G. Gerchikov and A.~V. Subashiev, Sov. Phys. Semicond. {\bf 26}, 73 (1992).

\bibitem{Hankiewicz06}
E.~M. Hankiewicz, G. Vignale, and M. Flatte, cond-mat/0603144.

\end{references}
\end{document}